\documentclass[aps,prl,preprint,tightenlines,superscriptaddress,showpacs,byrevtex]{revtex4}
\usepackage{graphicx} 
\usepackage{dcolumn}  

\graphicspath{{ps}}

\newcommand{\cp}{\ensuremath{B^{0} \rightarrow D^{*+} D^{*-} K^{0}_{S}}}
\newcommand{\cs}{\ensuremath{\bar B^{0} \rightarrow D^{*+} \bar D^{0} K^{-}}}
\newcommand{\res}{\ensuremath{B^{0} \rightarrow D^{+}_{{\rm s}1} \, (2536) D^{*-}}}

\newcommand{\epem}{\ensuremath{e^{+} e^{-}}}
\newcommand{\Ups}{\ensuremath{\Upsilon \, (4S)}}
\newcommand{\BBbar}{\ensuremath{B \bar B}}
\newcommand{\BzBzb}{\ensuremath{B^{0} - \bar B^{0}}}
\newcommand{\Bz}{\ensuremath{B^{0}}}
\newcommand{\Bzb}{\ensuremath{\bar B^{0}}}

\newcommand{\Bcp}{\ensuremath{B_{CP}}}
\newcommand{\Btag}{\ensuremath{B_{\rm Tag}}}

\newcommand{\DspKs}{\ensuremath{D^{*+} K^{0}_{\rm S}}}

\newcommand{\Dsone}{\ensuremath{D^{+}_{{\rm s}1} \, (2536)}}

\newcommand{\Dsp}{\ensuremath{D^{*+}}}
\newcommand{\Dsm}{\ensuremath{D^{*-}}}

\newcommand{\Dz}{\ensuremath{D^{0}}}

\newcommand{\Dp}{\ensuremath{D^{+}}}

\newcommand{\pip}{\ensuremath{\pi^{+}}}
\newcommand{\pim}{\ensuremath{\pi^{-}}}
\newcommand{\piz}{\ensuremath{\pi^{0}}}
\newcommand{\pips}{\ensuremath{\pi^{+}_{\rm slow}}}

\newcommand{\pizs}{\ensuremath{\pi^{0}_{\rm slow}}}
\newcommand{\Kp}{\ensuremath{K^{+}}}
\newcommand{\Km}{\ensuremath{K^{-}}}
\newcommand{\Ks}{\ensuremath{K^{0}_{S}}}
\newcommand{\Kl}{\ensuremath{K^{0}_{L}}}

\newcommand{\Mbc}{\ensuremath{M_{\rm bc}}}
\newcommand{\De}{\ensuremath{\Delta E}}
\newcommand{\MbcDe}{\ensuremath{M_{\rm bc} - \Delta E}}

\newcommand{\Dt}{\ensuremath{\Delta t}}
\newcommand{\Dmd}{\ensuremath{\Delta m_{\rm d}}}

\newcommand{\Acp}{\ensuremath{{\cal A}_{CP}}}
\newcommand{\Scp}{\ensuremath{D \sin 2 \phi_{1}}}
\newcommand{\Acpbe}{\ensuremath{\frac{J_{\rm c}}{J_{0}}}}
\newcommand{\Acpse}{\ensuremath{{J_{\rm c}}/{J_{0}}}}
\newcommand{\Scpbe}{\ensuremath{\frac{2 J_{{\rm s}1}}{J_{0}}\sin 2 \phi_{1}}}
\newcommand{\Scpse}{\ensuremath{{2 J_{{\rm s}1}}/{J_{0}}\sin 2 \phi_{1}}}

\newcommand{\Ccpb}{\ensuremath{\frac{2 J_{{\rm s}2}}{J_{0}}\cos 2 \phi_{1}}}
\newcommand{\Ccps}{\ensuremath{{2 J_{{\rm s}2}}/{J_{0}}\cos 2 \phi_{1}}}

\begin{document}

\vspace*{\baselineskip}
\resizebox{!}{3cm}{\includegraphics{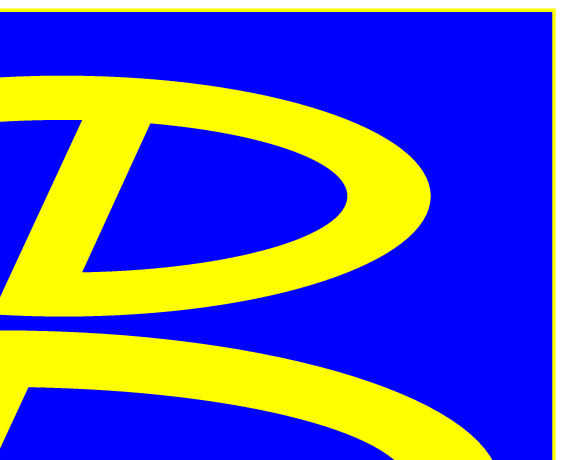}}

\preprint{\vbox{ \hbox{   }
    \hbox{}
    \hbox{}
    \hbox{}
    \hbox{}
    \hbox{}
    \hbox{}
    \hbox{BELLE Preprint 2007-25}
    \hbox{KEK Preprint 2007-14}
}}

\title{ \quad\\[1.0cm] Measurement of Branching Fraction and Time-Dependent $\mathbf{CP}$ Asymmetry Parameters in ${\mathbf\cp}$ Decays}

\begin{abstract}
  We present a measurement of the branching fraction and time-dependent $CP$ violation parameters for \cp\ decays. These results are obtained from a 414 ${\rm fb}^{-1}$ data sample that contains $449 \times 10^{6}$ \BBbar\ pairs collected at the \Ups\ resonance with the Belle detector at the KEKB asymmetric-energy \epem\ collider. We obtain the branching fraction,
  \begin{center}
    ${\cal B} \, (\cp) = [3.4 \pm 0.4 \; \rm{(stat)} \pm 0.7 \; \rm{(syst)}] \times 10^{-3}$,
  \end{center}
  which is in agreement with the current world average. We also obtain an upper limit on the product branching fraction for a possible two-body decay,
  \begin{center}
    ${\cal B} \, (\res){\cal B} \, (\Dsone \rightarrow \Dsp \; \Ks) < 7.1 \times 10^{-4}$ ($90\%$ CL).
  \end{center}
  In the traditional 2-parameter time-dependent $CP$ analysis, we measure the $CP$ violation parameters,
  \begin{center}
    \begin{tabular}{c}
      $\Acp = -0.01^{+0.28}_{-0.28} \; \rm{(stat)} \; \pm 0.09 \; \rm{(syst)}$\\
      $\Scp = 0.06^{+0.45}_{-0.44} \; \rm{(stat)} \; \pm 0.06 \; \rm{(syst)}$.\\
    \end{tabular}
  \end{center}
  No evidence for either mixing-induced or direct $CP$ violation is found. In a 3-parameter fit sensitive to $\cos 2 \phi_{1}$ performed in the half-Dalitz spaces, $s^{-} \leq s^{+}$ and $s^{-} > s^{+}$, where $s^{\pm} \equiv m^{2} \, (D^{*\pm}\Ks)$, we extract the $CP$ violation parameters,
  \begin{center}
    \begin{tabular}{c}
      $\Acpse = 0.60^{+0.25}_{-0.28} \; \rm{(stat)} \; \pm 0.08 \; \rm{(syst)}$\\
      $\Scpse = -0.17^{+0.42}_{-0.42} \; \rm{(stat)} \; \pm 0.09 \; \rm{(syst)}$\\
      $\Ccps = -0.23^{+0.43}_{-0.41} \; \rm{(stat)} \; \pm 0.13 \; \rm{(syst)}$.\\
    \end{tabular}
  \end{center}
  A large value of \Acpse\ would indicate a significant resonant contribution from a broad unknown $D^{**+}_{\rm s}$ state. Although the sign of the factor, ${2J_{{\rm s}2}}/{J_{0}}$, can be deduced from theory, no conclusion can be drawn regarding the sign of $\cos 2 \phi_{1}$ given the errors.
\end{abstract}


\affiliation{Budker Institute of Nuclear Physics, Novosibirsk}
\affiliation{Chiba University, Chiba}
\affiliation{University of Cincinnati, Cincinnati, Ohio 45221}
\affiliation{The Graduate University for Advanced Studies, Hayama}
\affiliation{Hanyang University, Seoul}
\affiliation{University of Hawaii, Honolulu, Hawaii 96822}
\affiliation{High Energy Accelerator Research Organization (KEK), Tsukuba}
\affiliation{University of Illinois at Urbana-Champaign, Urbana, Illinois 61801}
\affiliation{Institute of High Energy Physics, Vienna}
\affiliation{Institute of High Energy Physics, Protvino}
\affiliation{Institute for Theoretical and Experimental Physics, Moscow}
\affiliation{J. Stefan Institute, Ljubljana}
\affiliation{Kanagawa University, Yokohama}
\affiliation{Korea University, Seoul}
\affiliation{Kyungpook National University, Taegu}
\affiliation{Swiss Federal Institute of Technology of Lausanne, EPFL, Lausanne}
\affiliation{University of Maribor, Maribor}
\affiliation{University of Melbourne, School of Physics, Victoria 3010}
\affiliation{Nagoya University, Nagoya}
\affiliation{Nara Women's University, Nara}
\affiliation{National Central University, Chung-li}
\affiliation{National United University, Miao Li}
\affiliation{Department of Physics, National Taiwan University, Taipei}
\affiliation{H. Niewodniczanski Institute of Nuclear Physics, Krakow}
\affiliation{Nippon Dental University, Niigata}
\affiliation{Niigata University, Niigata}
\affiliation{University of Nova Gorica, Nova Gorica}
\affiliation{Osaka City University, Osaka}
\affiliation{Osaka University, Osaka}
\affiliation{Panjab University, Chandigarh}
\affiliation{Peking University, Beijing}
\affiliation{RIKEN BNL Research Center, Upton, New York 11973}
\affiliation{University of Science and Technology of China, Hefei}
\affiliation{Seoul National University, Seoul}
\affiliation{Sungkyunkwan University, Suwon}
\affiliation{University of Sydney, Sydney, New South Wales}
\affiliation{Tata Institute of Fundamental Research, Mumbai}
\affiliation{Toho University, Funabashi}
\affiliation{Tohoku University, Sendai}
\affiliation{Department of Physics, University of Tokyo, Tokyo}
\affiliation{Tokyo Institute of Technology, Tokyo}
\affiliation{Tokyo Metropolitan University, Tokyo}
\affiliation{Tokyo University of Agriculture and Technology, Tokyo}
\affiliation{Virginia Polytechnic Institute and State University, Blacksburg, Virginia 24061}
\affiliation{Yonsei University, Seoul}
  \author{J.~Dalseno}\affiliation{University of Melbourne, School of Physics, Victoria 3010}
  \author{I.~Adachi}\affiliation{High Energy Accelerator Research Organization (KEK), Tsukuba} 
  \author{H.~Aihara}\affiliation{Department of Physics, University of Tokyo, Tokyo} 
  \author{T.~Aushev}\affiliation{Swiss Federal Institute of Technology of Lausanne, EPFL, Lausanne}\affiliation{Institute for Theoretical and Experimental Physics, Moscow} 
  \author{A.~M.~Bakich}\affiliation{University of Sydney, Sydney, New South Wales} 
  \author{V.~Balagura}\affiliation{Institute for Theoretical and Experimental Physics, Moscow} 
  \author{A.~Bay}\affiliation{Swiss Federal Institute of Technology of Lausanne, EPFL, Lausanne} 
  \author{U.~Bitenc}\affiliation{J. Stefan Institute, Ljubljana} 
  \author{I.~Bizjak}\affiliation{J. Stefan Institute, Ljubljana} 
  \author{A.~Bozek}\affiliation{H. Niewodniczanski Institute of Nuclear Physics, Krakow} 
% ...
\collaboration{The Belle Collaboration}
\noaffiliation

\maketitle


{\renewcommand{\thefootnote}{\fnsymbol{footnote}}}
\setcounter{footnote}{0}

$CP$ violation in the Standard Model is due to a complex phase in the Cabibbo-Kobayashi-Maskawa (CKM) quark-mixing matrix~\cite{C,KM}. At present, $CP$ violation has been clearly observed by the BaBar~\cite{BABAR_jpsiks} and Belle~\cite{Belle_jpsiks} collaborations in the decay $B^{0} \rightarrow J/\psi \; K^{0}_{\rm S}$, while many other modes provide additional information on $CP$ violating parameters. One such decay is $\cp$, where $\sin 2 \phi_{1}$ can, in principle, be extracted from the time-dependent rate asymmetry~\cite{dspdsmks},
\begin{equation}
  \frac{\bar \Gamma \, (t) - \Gamma \, (t)}{\bar \Gamma \, (t) + \Gamma \, (t)} = \Acp \cos \, (\Dmd t) + \Scp \sin \, (\Dmd t ).
  \label{eq_ass_2par}
\end{equation}
$\Gamma \, (t)$ is the decay rate for \cp\ at proper time, $t$, after production while $\bar \Gamma \, (t)$ represents the charge conjugate rate. \Dmd\ is the mass difference between the two \Bz\ mass eigenstates, \Acp\ is the direct $CP$ violating component and $\sin 2 \phi_{1}$ measures mixing induced $CP$ violation. The factor, $D$, is the dilution of the $CP$ asymmetry and may arise from two sources. The contribution from different waves in the amplitude of \cp\ influences the $CP$ mixture and polarization of the final state. Furthermore, if an intermediate \DspKs\ resonance exists, additional dilution of the $CP$ asymmetry occurs as the decay becomes self-tagging. $CP$ contamination from penguins and final state interactions are expected to be small in this mode~\cite{dspdsmks}.

If resonant structure were to exist, $\cos 2 \phi_{1}$ may also be extracted, assuming no direct $CP$ violation, from the time-dependent rate asymmetry~\cite{dspdsmks},
\begin{equation}
  \frac{\bar \Gamma \, (t) - \Gamma \, (t)}{\bar \Gamma \, (t) + \Gamma \, (t)} = \eta_{y} \Acpbe \cos \, (\Dmd t) - \biggl(\Scpbe + \eta_{y} \Ccpb \biggr) \sin \, (\Dmd t ),
  \label{eq_ass_3par}
\end{equation}
in the half-Dalitz space, $s^{-} \leq s^{+}$ or $s^{-} > s^{+}$. The Dalitz variables are defined as $s^{\pm} \equiv m^{2} \, (D^{*\pm}\Ks)$ and $\eta_{y} = +1 \, (-1)$ if $s^{-} \leq s^{+} \, (s^{-} > s^{+}$). $J_{\rm c}$, $J_{0}$, $J_{{\rm s}1}$ and $J_{{\rm s}2}$ are integrals over the half-Dalitz space, $s^{-} > s^{+}$, of $|a|^{2} + |\bar a|^{2}$, $|a|^{2} - |\bar a|^{2}$, the real component, $\Re \, (\bar a a^{*})$ and the imaginary component, $\Im \, (\bar a a^{*})$, respectively, where $a \, (\bar a)$ are the decay amplitudes of $\Bz \, (\Bzb) \rightarrow \Dsp \Dsm \Ks$. Note that $J_{{\rm s}2}$ is non-zero only if \cp\ has a resonant component: in this case, $J_{\rm c}$ may be large.

This measurement of the branching fraction and $CP$ violating parameters in $\cp$ is based on a $414~{\rm fb}^{-1}$ data sample that contains $449 \times 10^6 \; B \bar B$ pairs, collected  with the Belle detector at the KEKB asymmetric-energy $e^+e^-$ ($3.5$ on $8~{\rm GeV}$) collider~\cite{KEKB}. Operating with a peak luminosity that exceeds $1.6\times 10^{34}~{\rm cm}^{-2}{\rm s}^{-1}$, the collider produces the \Ups\ resonance ($\sqrt{s}=10.58$~GeV) with a Lorentz boost of $\beta\gamma=0.425$, opposite to the positron beamline direction, $z$. Since the $B^0$ and $\bar{B}^0$ mesons are approximately at rest in the \Ups\ Center-of-Mass System (CMS), the difference in decay time between the $B \bar B$ pair, $\Delta t$, can be determined from the displacement in $z$ between the final state decay vertices,
\begin{equation}
  \Dt \simeq \frac{(z_{CP} - z_{\rm Tag})}{\beta \gamma c} \equiv \frac{\Delta z}{\beta \gamma c}.
  \label{delta_t}
\end{equation}
The subscripts, $CP$ and Tag, denote the reconstructed final state, \Bcp, and the final state from which the flavor is identified, \Btag. The quantity, \Dt, can be substituted for $t$ in (\ref{eq_ass_2par}) and (\ref{eq_ass_3par}) due to coherent \BBbar\ production at the \Ups\ resonance~\cite{Sanda}.

The Belle detector is a large-solid-angle magnetic
spectrometer that
consists of a silicon vertex detector (SVD),
a 50-layer central drift chamber (CDC), an array of
aerogel threshold Cherenkov counters (ACC), 
a barrel-like arrangement of time-of-flight
scintillation counters (TOF), and an electromagnetic calorimeter
comprised of CsI (Tl) crystals (ECL) located inside 
a superconducting solenoid coil that provides a 1.5~T
magnetic field.  An iron flux-return located outside of
the coil is instrumented to detect \Kl\ mesons and to identify
muons (KLM).  The detector
is described in detail elsewhere~\cite{Belle}.
Two inner detector configurations were used. A 2.0 cm beampipe
and a 3-layer silicon vertex detector (SVD1) was used for the first sample
of 152 $\times 10^6 B\bar{B}$ pairs, while a 1.5 cm beampipe, a 4-layer
silicon detector (SVD2) and a small-cell inner drift chamber was used to record  
the remaining 297 $\times 10^6 B\bar{B}$ pairs~\cite{Ushiroda}.  

We reconstruct \cp\ from $\Dsp \rightarrow \Dz \pips$ and $\Dp \pizs$ requiring at least one $\Dz \pips$ decay. \Dz\ and \Dp\ candidates are reconstructed from $\Dz \rightarrow \Km \pip$, $\Ks \pip \pim$, $\Km \pip \piz$, $\Km \pip \pip \pim$, $\Km \Kp$ and $\Dp \rightarrow \Km \pip \pip$, $\Km \Kp\ \pip$; cases with two $\Dz \rightarrow \Ks \pip \pim$ decays are rejected. We also reconstruct $\Ks \rightarrow \pip \pim$ and $\piz \rightarrow \gamma \gamma$. The reconstruction of the charge conjugates in the above decay chain is implied.

According to Monte Carlo (MC) studies, continuum background does not give a significant contribution to this mode, so only a loose requirement on the ratio of the second to zeroth Fox-Wolfram moment~\cite{SFW} is used, $R_{2} < 0.4$. Charged tracks satisfy loose criteria on their impact parameters relative to the interaction point (IP), $|dr| < 0.4 \; {\rm cm}$ and $|dz| < 5.0 \; {\rm cm}$ with some additional SVD requirements~\cite{res_func}. With information obtained from the CDC, ACC and TOF, particle identification (PID), or $K / \pi$ separation is determined with the likelihood ratio, ${\cal L}_{K}/({\cal L}_{K} + {\cal L}_{\pi})$. Here, ${\cal L}_{i}$ is the likelihood that the particle is of type $i$. Candidate \piz's are selected from pairs of photons with energies greater than $30$ MeV. The \piz\ momentum is required to be greater than $200$ MeV/$c$ in the laboratory frame. Charged slow pions used in the reconstruction of $D^{*}$'s, are not subject to impact parameter or PID cuts. Similarly, the \piz\ momentum cut is not applied to slow \piz's used to reconstruct $D^{*}$'s.

We generate signal MC with EvtGen~\cite{EvtGen} using a detector simulation based on GEANT3~\cite{GEANT}. All mass windows are chosen from correctly reconstructed signal MC by fitting the mass distributions with double Gaussians and applying a $3 \sigma$ cut on the wider Gaussian. This method is applied to \piz, \Ks, $D$ and the $D^{*} - D$ mass difference for $D^{*}$ selection. When multiple $B$ mesons are reconstructed, the candidate with the smallest value of
\begin{equation}
  \chi^{2}_{\rm mass} \equiv \sum_{i} \chi^{2} \, (X), \;\; {\rm where} \; \chi^{2} \, (X) \equiv \frac{|{\rm mass} \, (X) - {\rm mass} \, (X)_{\rm PDG}|^{2}}{\sigma \, (X)^{2}},
\end{equation}
is selected. We sum over the prompt \Ks, the two daughter $D$ mesons and the two daughter $D^{*}$'s where $\sigma \, (X)$ is the width of the narrow Gaussian found in the fit to the invariant mass of particle $X$.

As we reconstruct the decay chain, vertexing algorithms are applied to \Ks\ and $D$ mesons from their charged daughters. The vertex of the $B$ is determined from the pseudo-tracks of the two daughter $D$ mesons with an IP constraint. We do not include the slow pions in the vertex fit due to their poor resolution. The prompt \Ks\ pseudo-track is also excluded as its decay vertex has a relatively large displacement from the IP. However, both the slow pion and \Ks\ candidate are constrained to originate from the $B$ vertex determined by the $D^{*}$ mesons and their momentum is recalculated to improve the resolution of the $B$ candidate.

\begin{figure}[htb]
  \centering
  \includegraphics[height=250pt,width=!]{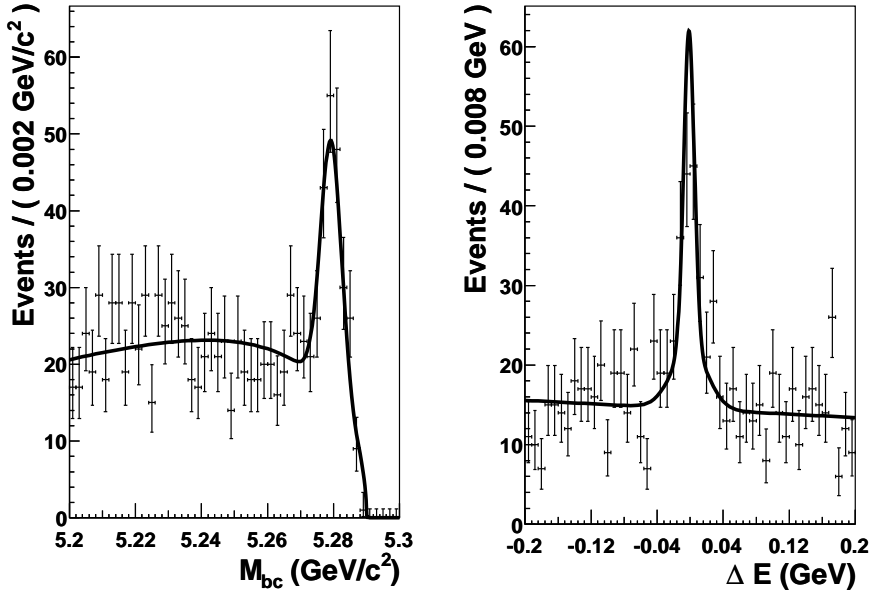}
  \caption{\MbcDe\ fit projections. The solid curve shows the fit result.}
  \label{fig_mbcde}
\end{figure}

The $B$ decay is described by the variables $\Mbc \equiv \sqrt{(E^{\rm CMS}_{\rm beam})^{2} - (p^{\rm CMS}_{B})^{2}}$ and $\De \equiv E^{\rm CMS}_{B} - E^{\rm CMS}_{\rm beam}$. The \Mbc\ signal shape is modeled by a Gaussian while the background is modeled with an ARGUS function~\cite{ARGUS}. In $\Delta E$, the signal shape is represented by a double Gaussian and the background with a linear function. With the signal shape determined from MC, we define the signal box to be $3 \sigma$ of the Gaussian in \Mbc\ by $3 \sigma$ of the narrow Gaussian in \De, which is $5.27$ GeV/$c^{2}$ $< \Mbc < 5.29$ GeV/$c^{2}$ and $-0.04$ GeV $< \De < 0.04$ GeV. The reconstruction efficiency, which includes all intermediate branching fractions, is determined from phase space signal MC to be $(6.59 \pm 0.13) \times 10^{-5}$ for SVD1 and $(9.58 \pm 0.16) \times 10^{-5}$ for SVD2, giving an average efficiency of $(8.57 \pm 0.15) \times 10^{-5}$. The noticeable difference in efficiency between SVD1 and SVD2 is due to the better performance of SVD2 with low momentum tracking. With the signal shape fixed from signal MC, we perform an extended two dimensional maximum likelihood fit to obtain the yield (Fig.~\ref{fig_mbcde}), ${\rm Y} = 131.2^{+14.8}_{-14.1} \; {\rm (stat)}$ events. Using
\begin{equation}
  {\cal B} \, (\cp) = \frac{{\rm Y}}{\epsilon \, {\rm N} \, (B \bar B)},
\end{equation}
where $\epsilon$ is the average detection efficiency and ${\rm N} \, (B \bar B)$ is the number of \BBbar\ pairs, we obtain the branching fraction,
\begin{equation}
  {\cal B} \, (\cp) = [3.4 \pm 0.4 \; {\rm (stat)} \pm 0.7 \; {\rm (syst)}] \times 10^{-3},
  \label{eq_bf}
\end{equation}
which relies on the assumption of equal production of neutral and charged $B$ meson pairs from the \Ups.

Systematic errors are estimated from various sources such as the number of \BBbar\ events, the uncertainty in the efficiency and the PDG errors on branching fractions~\cite{PDG}. The \Ks, \piz, PID and tracking systematics are determined from the efficiency ratio of data to MC as calculated by independent studies at Belle. We also include the difference in yield between fits with signal shape parameters obtained directly from data, and obtained from MC, as a systematic error. The wider $\Delta E$ Gaussian in the fit to data is fixed, from MC, relative to the narrow floating Gaussian. Peaking background due to fake \pizs's is also accounted for by investigating the \Dsp\ sideband. We also include possible efficiency variations across the Dalitz plane by comparing the average efficiency of phase space signal MC with the average efficiency as determined from bins of $m \, (D^{*\pm} \Ks)$ using data weights. The systematics are summarized in Table~\ref{tab_sys_bf}.
\begin{table}[htb]
  \begin{tabular}
    {@{\hspace{0.5cm}}l@{\hspace{0.5cm}}||@{\hspace{0.5cm}}c@{\hspace{0.5cm}}}
    \hline \hline
    Source & Systematic error (\%)\\
    \hline
    Number of \BBbar\ pairs & $1.3$\\
    MC reconstruction efficiency & $1.7$\\
    \Ks, \piz and PID efficiency & $5.8$\\
    Daughter branching fractions & $8.4$\\
    Tracking efficiency & $11.9$\\
    Signal shape & $10.4$\\
    \piz\ peaking background & $4.7$\\
    Dalitz efficiency dependence & $2.6$\\
    \hline
    Total & $19.8$\\
    \hline \hline
  \end{tabular}
  \caption{\cp\ branching fraction systematics.}
  \label{tab_sys_bf}
\end{table}

\begin{figure}[htb]
  \centering
  \includegraphics[height=165pt,width=!]{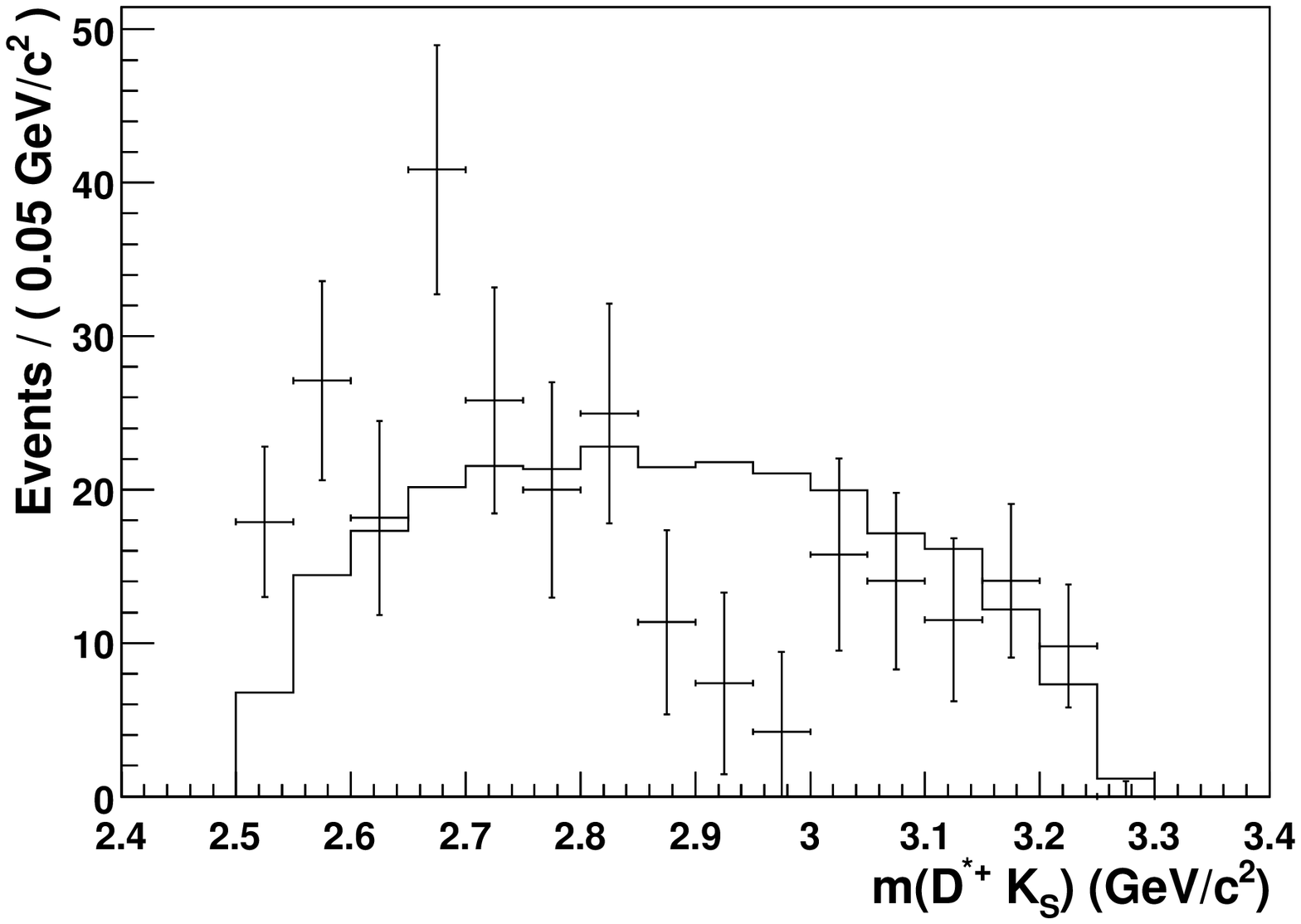}
  \includegraphics[height=165pt,width=!]{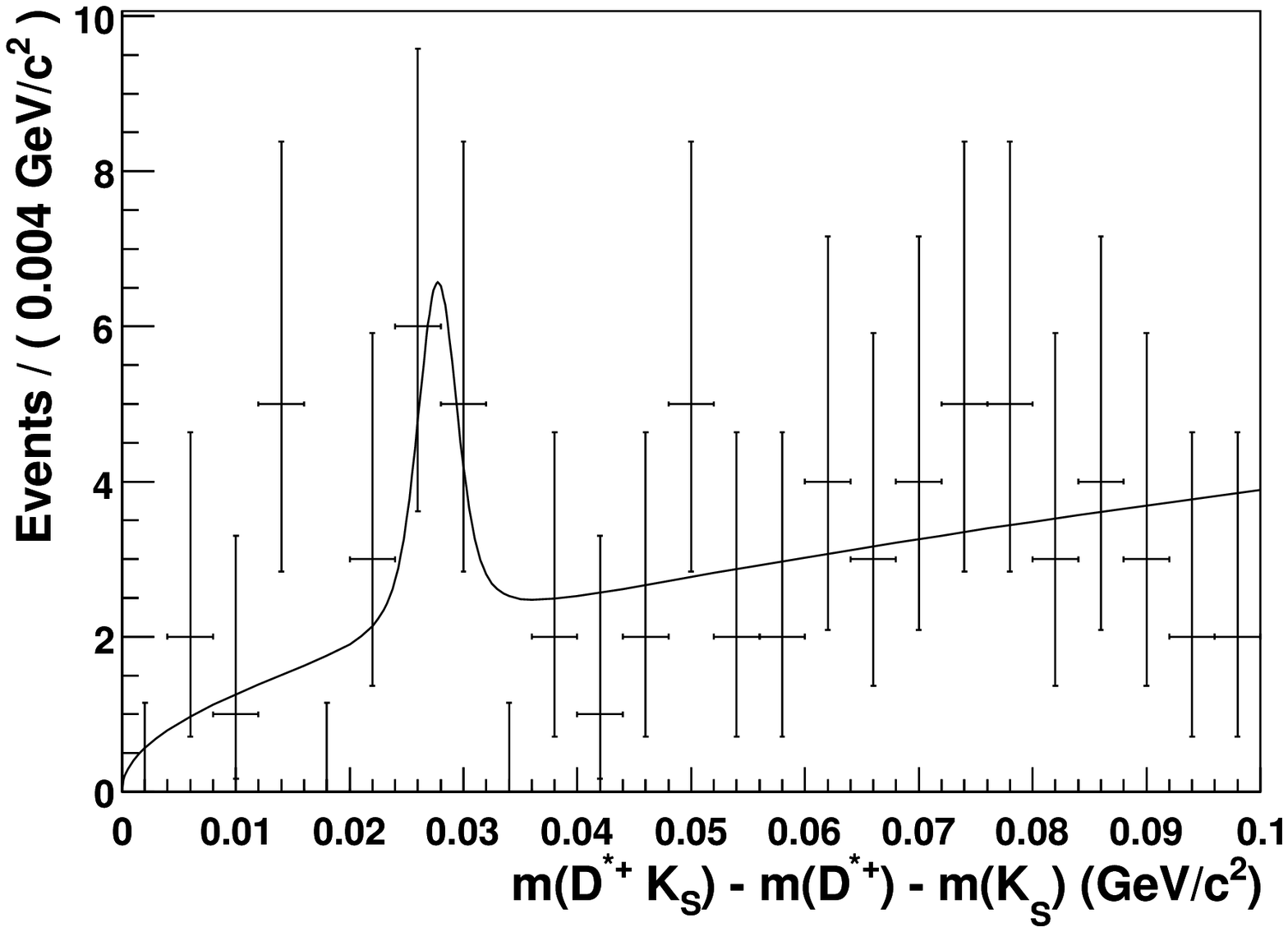}
  \put(-274,135){(a)}
  \put(-42,135){(b)}
  \caption{(a) Background subtracted \DspKs\ raw yield vs. invariant mass with a superimposed reconstructed phase space histogram from signal MC. (b) Measured distribution of $m \, (\DspKs) - m \, (\Dsp) - m \, (\Ks)$ in the signal box. The solid line shows the fit results.}
  \label{fig_resonance}
\end{figure}

The $D^{*\pm} \Ks$ spectrum inside the signal box is examined to search for resonant structure. Fig.~\ref{fig_resonance}(a) shows the background subtracted $D^{*\pm} \Ks$ raw yield versus invariant mass overlaid with reconstructed phase space obtained from signal MC. The background is taken from a sideband defined to be $5.20$ GeV/$c^{2}$ $< \Mbc < 5.26$ GeV/$c^{2}$ and $-0.2$ GeV $< \De < 0.2$ GeV. From this plot, the data inside the signal box does not appear to be entirely consistent with phase space, but neither is there any easily identifiable peak.

We extract the signal yield of \res\ from the mass difference, $\Delta m = m \, (\DspKs) - m \, (\Dsp) - m \, (\Ks)$, in the signal box. Note that while this form of $\Delta m$ has been chosen, the \Ks\ was fitted to its nominal mass. A Breit-Wigner function convolved with a Gaussian is chosen to model the signal shape, while a simple threshold function, $P_{\rm Bkg} \, (\Delta m) = c \Delta m^{1/2}$, is chosen to describe the background, where $c$ is a normalizing constant. In the fit to signal MC, the Breit-Wigner mass and width are fixed to their respective generated values while the Gaussian is floated to determine the effects of detector smearing. In the fit to data, the same smearing Gaussian is used, but the Breit-Wigner width is conservatively increased to the current upper limit on the \Dsone\ width, 2.3 MeV/$c^{2}$~\cite{PDG}. The fitted yield is ${\rm Y} \, (\res) = 6.2^{+4.0}_{-3.4} \; {\rm (stat)}$ events, and is shown in Fig.~\ref{fig_resonance}(b). The detection efficiency in this region is found to be $(3.08 \pm 0.40) \times 10^{-5}$ for SVD1 and $(3.98 \pm 0.36) \times 10^{-5}$ for SVD2, giving an average efficiency of $(3.68 \pm 0.38) \times 10^{-5}$, and we obtain the product of branching fractions,
\begin{equation}
  {\cal B} \, (\res){\cal B} \, (\Dsone \rightarrow \Dsp \; \Ks) = [3.8^{+2.4}_{-2.1} \; {\rm (stat)} \; ^{+0.8}_{-0.7} \; {\rm (syst)}] \times 10^{-4},
\end{equation}
where the systematics were determined in the same way as for \cp. Thus, the upper limit is found to be
\begin{equation}
  {\cal B} \, (\res){\cal B} \, (\Dsone \rightarrow \Dsp \; \Ks) < 7.1 \times 10^{-4} \;\; (90\% {\rm \; CL}).
  \label{eq_bf_ds1pdsm}
\end{equation}

To obtain the \Dt\ distribution, we reconstruct the tag-side vertex from the tracks not used to reconstruct the $CP$-side~\cite{res_func}, and employ the flavor tagging routine described in Ref.~\cite{TaggingNIM}. Due to imperfect flavor tagging, both a flavor tag, $q$, and the dilution factor of flavor tagging, $r$, are assigned to each event. The $r$ distribution is divided into 6 bins.

In the 2-parameter fit, the time-dependent $\Delta t$ distribution of $\cp$ is given by the probability density function (PDF),
\begin{equation}
  {\cal P}_{\rm Sig} \, (\Dt, q) \equiv \frac{e^{-|\Dt|/\tau_{\Bz}} }{4\tau_{\Bz}} \biggl\{ 1 - q \Delta w + q \, (1 - 2w)
  [ \Acp \cos \, (\Dmd \Dt) + \Scp \sin \, (\Dmd \Dt) ] \; \biggr\},
  \label{sig_tcpv_pdf}
\end{equation}
where $q$ is the tagged flavor, and $w$, $\Delta w$ account for imperfect flavor tagging. $w$ is the probability of mis-tagging an event, $\Delta w$ is the difference between $B^{0}$ and $\bar B^{0}$ in $w$ and they are both determined from flavor specific control samples~\cite{TaggingNIM}. \Acp\ and \Scp\ are the two free parameters in this fit. To account for differences between true $\Delta t$ and reconstructed $\Delta t$, ${\cal P}_{\rm Sig}$ is convolved with a resolution function described in Ref.~\cite{res_func},
\begin{equation}
  P_{\rm Sig} \, (\Dt, q) = \int_{-\infty}^{+\infty} {\cal P}_{\rm Sig} \, (\Dt', q) R_{\rm Sig} \, (\Dt - \Dt')d(\Dt'),
  \label{sig_tcpv_eq}
\end{equation}
which describes the effects of detector resolution and the effect of charm decays on the tag-side. The effect of the approximation in Eq. \ref{delta_t} is also included in the resolution function. The background PDF is modeled by an exponential lifetime component and a prompt component given by a delta function,
\begin{equation}
  {\cal P}_{\rm Bkg} \, (\Delta t) \equiv (1 - f_{\delta}) \frac{e^{-|\Delta t|/\tau_{\rm Bkg}} }{2\tau_{\rm Bkg}} + f_{\delta} \; \delta \, ( \Delta t - \mu_{\delta}),
\end{equation}
where $\tau_{\rm Bkg}$ is the effective lifetime of the background and $\mu_{\delta}$ sets the $\Delta t$ location of the prompt component. The PDF, ${\cal P}_{\rm Bkg}$, is convolved with a resolution function that takes the form of a double Gaussian, to give the background PDF, $P_{\rm Bkg}$. The background function parameters are determined from the sideband data defined by $5.20$ GeV/$c^{2}$ $< M_{\rm bc} < 5.26$ GeV/$c^{2}$ and $-0.2$ GeV $< \De < 0.2$ GeV. To account for events with large $|\Dt|$ that are not modeled by either the signal or background PDFs, an outlier background Gaussian, $P_{\rm Ol}$, is also included. The probability that an event is a signal event is assigned from the \MbcDe\ PDF based on the event's position inside the signal box,
\begin{equation}
  f_{\rm Sig} \, (\Mbc, \De, r) \equiv \frac{p \, (r) {\cal P}_{\rm Sig} \, (\Mbc, \De)}{p \, (r) {\cal P}_{\rm Sig} \, (\Mbc, \De) + (1-p \, (r)) {\cal P}_{\rm Bkg} \, (\Mbc, \De)},
  \label{sig_frac_tcpv_eq}
\end{equation}
where $p \, (r)$ is the purity in the signal region of each $r$-bin. Finally, the full time-dependent $CP$ violation PDF is written as,
\begin{eqnarray}
  P \, (\Dt, q, \Mbc, \De, r) = (1-f_{\rm Ol}) \biggl\{ && \!\!\!\!\!\! f_{\rm Sig} \, (\Mbc, \De, r) P_{\rm Sig} \, (\Dt, q) +\nonumber \\
    && \!\!\!\!\!\! \frac{1}{2}[1-f_{\rm Sig} \, (\Mbc, \De, r)]P_{\rm Bkg} \, (\Dt) \biggr\} + \frac{1}{2}f_{\rm Ol}P_{\rm Ol} \, (\Dt).\nonumber \\
\end{eqnarray}

\begin{figure}[htb]
  \centering
  \includegraphics[height=230pt,width=!]{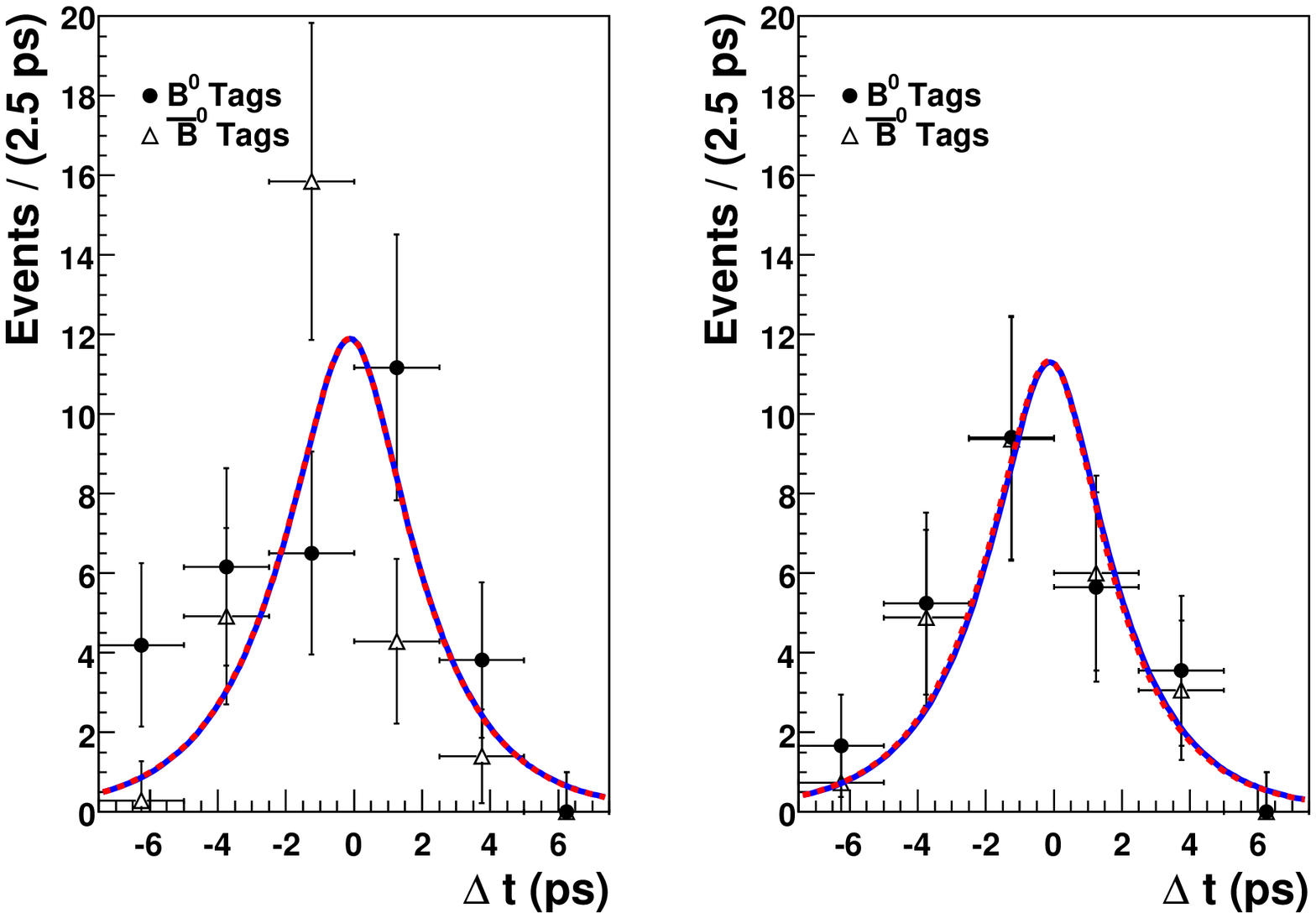}
  \put(-37,193){(b)}
  \put(-195,193){(a)}

  \includegraphics[height=230pt,width=!]{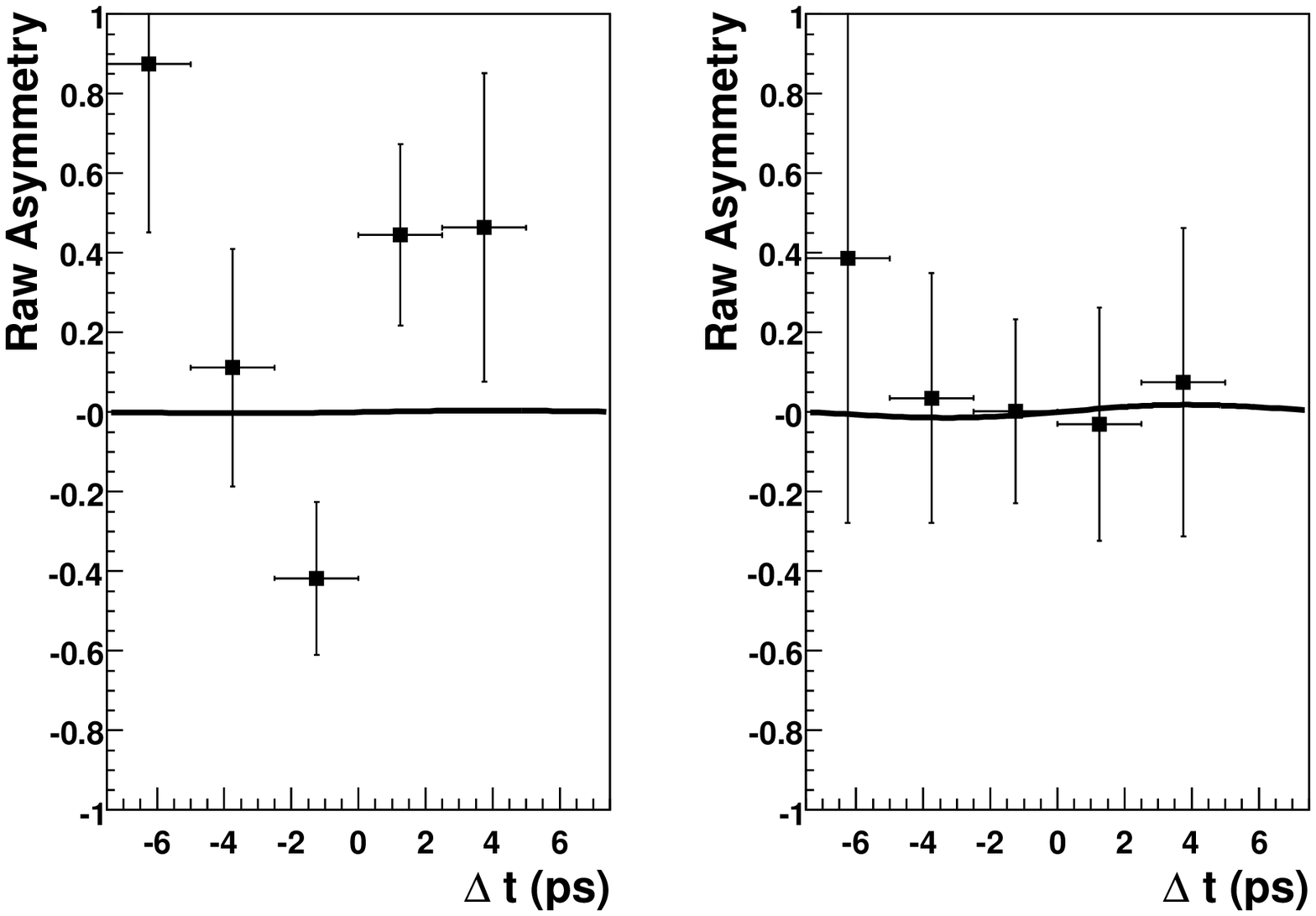}
  \put(-37,193){(d)}
  \put(-195,193){(c)}
  \caption{2-parameter fit results of \cp. (a) and (b) show the background subtracted \Dt\ distribution for poor tags ($0.0 < r \leq 0.5$) and good tags ($0.5 < r \leq 1.0$). Background subtraction is performed by subtracting the background PDF from the raw \Dt\ distribution. The solid curve represents \Bz\ tagged candidates and the dashed curve represents \Bzb\ tagged candidates. (c) and (d) show the \BzBzb\ raw asymmetry, $(N_{\Bz} - N_{\Bzb})/(N_{\Bz} + N_{\Bzb})$, where $N_{\Bz} \, (N_{\Bzb})$ is the number of \Bz \, (\Bzb) tags in \Dt\ for poor tags and good tags, respectively.}
  \label{fig_cp_2par}
\end{figure}

A control sample, \cs, is selected for which no $CP$ violation is expected. For this sample, the time-dependent fit reveals no $CP$ violation,
\begin{equation}
  \begin{tabular}{c}
    $\Acp \, (\cs) = 0.09 \pm 0.13 \; {\rm (stat)}$\\
    ${\cal S}_{CP} \, (\cs) = 0.01 \pm 0.20 \; {\rm (stat)}$,\\
  \end{tabular}
\end{equation}
where ${\cal S}_{CP}$ represents the component of mixing induced $CP$ violation. We then determine the $CP$ asymmetry parameters of \cp,
\begin{equation}
  \begin{tabular}{c}
    ${\cal A}_{CP} = -0.01^{+0.28}_{-0.28} \; ({\rm stat}) \pm 0.09 \; ({\rm syst})$\\
    $D \sin 2 \phi_{1} = 0.06^{+0.45}_{-0.44} \; ({\rm stat}) \pm 0.06 \; ({\rm syst})$,\\
  \end{tabular}
  \label{eq_2par_fit}
\end{equation}
and the fit results are shown in Fig.~\ref{fig_cp_2par}.

The sources of systematic errors are addressed as follows. The uncertainties in the vertex reconstruction are the dominant sources of error. These include uncertainties in the IP profile, charged track selection based on track helix errors, helix parameter corrections, \Dt\ and vertex goodness-of-fit selection, and SVD misalignment. Toy MC samples showed some small fitting bias for $CP$ parameters due to low statistics in each sample. We take this bias as a systematic. The parameters for $w$ and $\Delta w$, resolution function, physics parameters, background shape and signal fraction were varied by $\pm 1 \sigma$. We estimate that the $CP$ eigenstate component of the background does not exceed $10\%$ of the total. The effect of $CP$ asymmetry in the background was estimated by changing the background PDF to be composed of 10\% $CP$ eigenstates with $CP$ parameters at the physical limits. We also investigate the effect of mis-reconstructed signal events using signal MC. This is achieved by comparing the fit result of the signal MC sample with another fit on the same sample, which required that the events were reconstructed correctly. Tag-side interference comes from $CP$ violation on the tag-side~\cite{TSI}, and is estimated with $B \rightarrow D^{*} l \nu$. We generate MC pseudo-experiments and perform an ensemble test to obtain systematic biases~\cite{TSI2}. The systematics are summarized in Table~\ref{tcpv_syst}.
\begin{table}[htb]
  \begin{tabular}
    {@{\hspace{0.5cm}}l@{\hspace{0.5cm}}||@{\hspace{0.5cm}}c@{\hspace{0.5cm}}@{\hspace{0.5cm}}c@{\hspace{0.5cm}}}
    \hline \hline
    Source & $\delta(\Acp)$ & $\delta(\Scp)$\\
    \hline
    Vertex reconstruction & $0.064$ & $0.021$\\
    Wrong tag fraction & $0.020$ & $0.030$\\
    Resolution function & $0.019$ & $0.008$\\
    Fit bias & $0.0003$ & $0.003$\\
    Physics parameters & $0.001$ & $0.001$\\
    Background $\Delta t$ & $0.016$ & $0.019$\\
    Signal fraction & $0.005$ & $0.004$\\
    Mis-reconstructed events & $0.014$ & $0.038$\\
    Tag-side interference & $0.053$ & $0.001$\\
    \hline
    Total & $0.090$ & $0.058$\\
    \hline \hline
  \end{tabular}
  \caption{Time-dependent $CP$ asymmetry systematics: 2-parameter fit.}
  \label{tcpv_syst}
\end{table}

In the 3-parameter fit, the time-dependent $\Delta t$ distribution assuming no direct $CP$ violation, is given by
\begin{eqnarray}
  {\cal P}_{\rm Sig} \, (\Dt, q) &\equiv& \frac{e^{-|\Dt|/\tau_{\Bz}}}{4\tau_{\Bz}} \biggl\{ 1 - q \Delta w + q \, (1 - 2w) \times \nonumber \\
  && \biggl[ \eta_{y} \Acpbe \cos \, (\Dmd \Dt) - \biggl(\Scpbe + \eta_{y} \Ccpb \biggr) \sin \, (\Dmd \Dt ) \biggr] \; \biggr\}. \nonumber \\
\end{eqnarray}

\begin{figure}[htb]
  \centering
  \includegraphics[height=230pt,width=!]{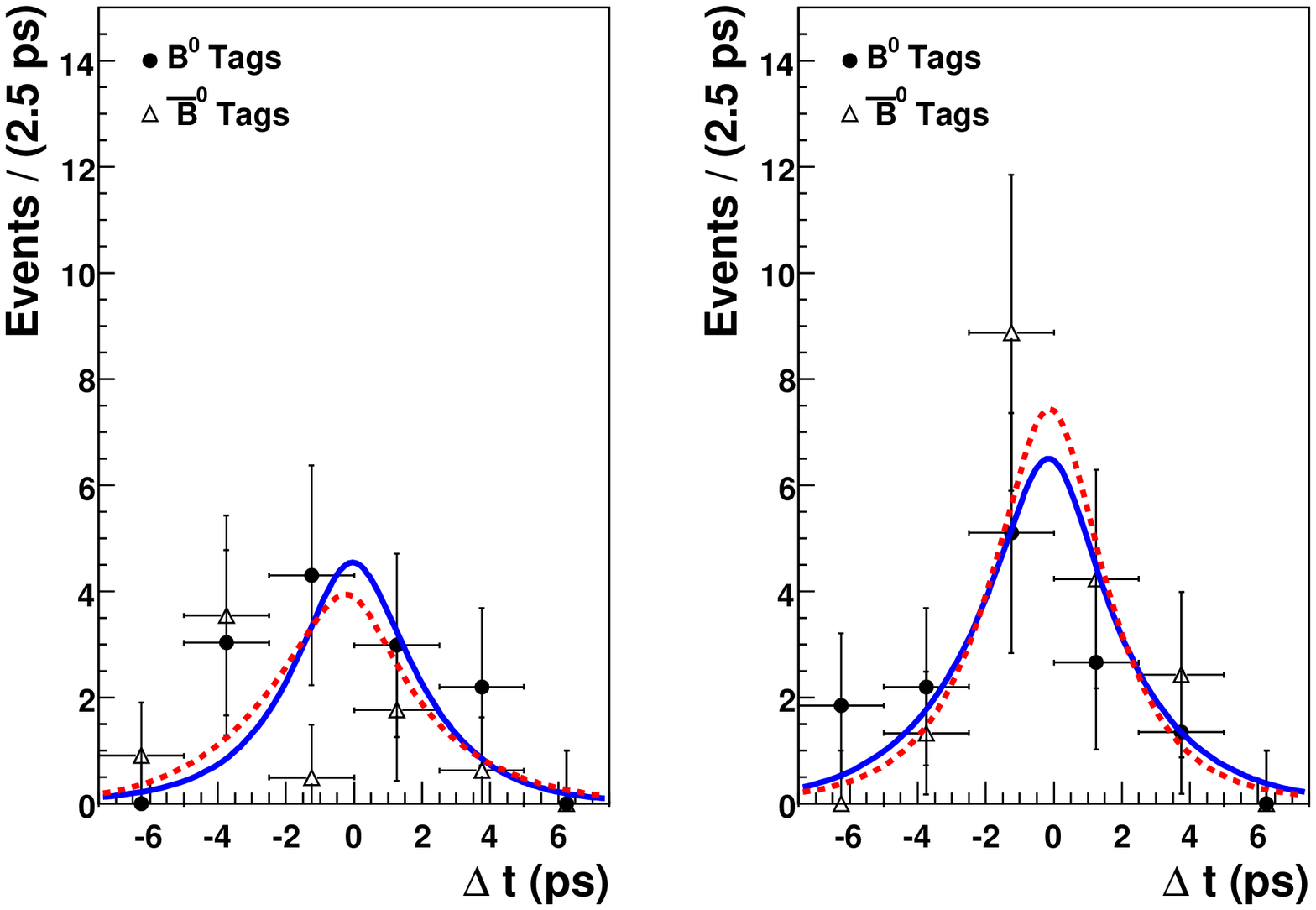}
  \put(-37,193){(b)}
  \put(-195,193){(a)}

  \includegraphics[height=230pt,width=!]{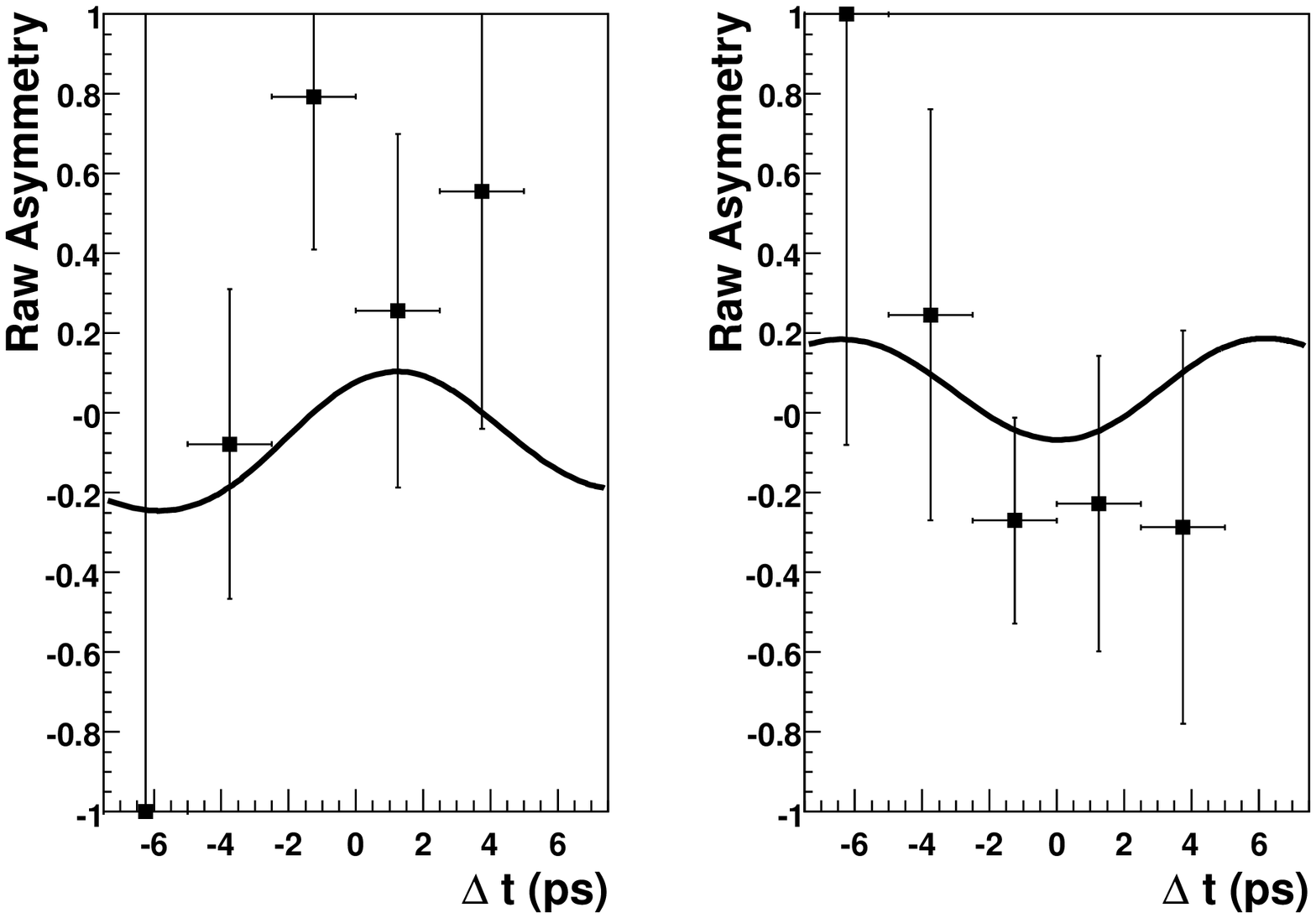}
  \put(-37,193){(d)}
  \put(-195,193){(c)}
  \caption{3-parameter fit results of \cp. (a) and (b) show the background subtracted \Dt\ distribution for good tags ($0.5 < r \leq 1.0$) in the region, $s^{-} \leq s^{+}$ and $s^{-} > s^{+}$. The solid curve represents \Bz\ tags and the dashed curve represents \Bzb\ tags. (c) and (d) show the \BzBzb\ raw asymmetry in \Dt\ for each half-Dalitz space and for good tags only.}
  \label{fig_cp_3par}
\end{figure}

We obtain the time-dependent $CP$ parameters,
\begin{equation}
  \begin{tabular}{c}
    $\Acpse = 0.60^{+0.25}_{-0.28} \; \rm{(stat)} \pm 0.08 \; \rm{(syst)}$\\
    $\Scpse = -0.17^{+0.42}_{-0.42} \rm{(stat)} \; \pm 0.09 \; \rm{(syst)}$\\
    $\Ccps = -0.23^{+0.43}_{-0.41} \rm{(stat)} \; \pm 0.13 \; \rm{(syst)}$,\\
  \end{tabular}
  \label{eq_3par_fit}
\end{equation}
and the corresponding fit curves are shown in Fig.~\ref{fig_cp_3par}.

The systematics are determined in much the same way as in the 2-parameter fit. However, since BaBar claims evidence for a \res\ contribution to the $\Dsp \; \Dsm \; \Ks$ final state~\cite{BABAR_dspdsmks}, and resonant structure is more important in the 3-parameter fit, we excluded the \Dsone\ region, $m \, (D^{*\pm}\Ks) < 2.6$ GeV/$c^{2}$, and took the difference in the results as a systematic. The systematics are summarized in Table~\ref{tcpv-3_syst}.

\begin{table}[htb]
  \begin{tabular}
    {@{\hspace{0.5cm}}l@{\hspace{0.5cm}}||@{\hspace{0.5cm}}c@{\hspace{0.5cm}}@{\hspace{0.5cm}}c@{\hspace{0.5cm}}@{\hspace{0.5cm}}c@{\hspace{0.5cm}}}
    \hline \hline
    Source & $\delta(\Acpse)$ & $\delta(\Scpse)$ & $\delta(\Ccps)$\\
    \hline
    Vertex reconstruction & $0.066$ & $0.071$ & $0.081$\\
    Wrong tag fraction & $0.021$ & $0.033$ & $0.032$\\
    Resolution function & $0.014$ & $0.039$ & $0.045$\\
    Fit bias & $0.029$ & $0.018$ & $0.031$\\
    Physics parameters & $0.005$ & $0.004$ & $0.003$\\
    Background $\Delta t$ & $0.005$ & $0.019$ & $0.012$\\
    Signal fraction & $0.020$ & $0.009$ & $0.008$\\
    Mis-reconstructed events & $0.020$ & $0.007$ & $0.026$\\
    Tag-side interference & $0.000$ & $0.002$ & $0.004$\\
    \Dsone & $0.019$ & $0.013$ & $0.069$\\
    \hline
    Total & $0.084$ & $0.094$ & $0.127$\\
    \hline \hline
  \end{tabular}
  \caption{Time-dependent $CP$ asymmetry systematics: 3-parameter fit.}
  \label{tcpv-3_syst}
\end{table}

In summary, we measure the branching fraction (Eq.~\ref{eq_bf}) and $CP$ parameters of \cp\ with a 2-parameter fit (Eq.~\ref{eq_2par_fit}), and 3-parameter fit (Eq.~\ref{eq_3par_fit}) with $414 \; {\rm fb}^{-1}$ at the Belle detector. An upper limit on the branching fraction of the intermediate two-body decay \res (Eq.~\ref{eq_bf_ds1pdsm}) is also obtained. In the 2-parameter fit, there is no evidence for direct $CP$ violation and a large dilution of $\sin 2 \phi_{1}$ is implied. In the 3-parameter fit, ${J_{\rm c}}/{J_{0}}$ appears to be non-zero; a large value of this parameter could indicate the presence of a broad unknown $D^{**+}_{\rm s}$ state. As in the 2-parameter fit, a large dilution of the $CP$ asymmetry from polarization and resonant structure is implied since ${2J_{{\rm s}1}}/{J_{0}} \sin 2 \phi_{1}$ is consistent with zero. According to~\cite{dspdsmks}, ${2J_{{\rm s}2}}/{J_{0}}$, is expected to be positive if this unknown wide resonance exists. The parameter, ${2J_{{\rm s}2}}/{J_{0}} \cos 2 \phi_{1}$, has been measured, however, a model-dependent inference on the sign of $\cos 2 \phi_{1}$ is not possible with current precision.

We thank the KEKB group for the excellent operation of the
accelerator, the KEK cryogenics group for the efficient
operation of the solenoid, and the KEK computer group and
the National Institute of Informatics for valuable computing
and Super-SINET network support. We acknowledge support from
the Ministry of Education, Culture, Sports, Science, and
Technology of Japan and the Japan Society for the Promotion
of Science; the Australian Research Council and the
Australian Department of Education, Science and Training;
the National Science Foundation of China and the Knowledge
Innovation Program of the Chinese Academy of Sciences under
contract No.~10575109 and IHEP-U-503; the Department of
Science and Technology of India; 
the BK21 program of the Ministry of Education of Korea, 
the CHEP SRC program and Basic Research program 
(grant No.~R01-2005-000-10089-0) of the Korea Science and
Engineering Foundation, and the Pure Basic Research Group 
program of the Korea Research Foundation; 
the Polish State Committee for Scientific Research; 
the Ministry of Education and Science of the Russian
Federation and the Russian Federal Agency for Atomic Energy;
the Slovenian Research Agency;  the Swiss
National Science Foundation; the National Science Council
and the Ministry of Education of Taiwan; and the U.S.\
Department of Energy.


\begin{thebibliography}{99}

\bibitem{C}
  N.~Cabibbo, Phys. Rev. Lett. {\bf 8}, 214 (1964).

\bibitem{KM}
  M.~Kobayashi and T.~Maskawa, Prog. Theor. Phys. {\bf 49}, 652 (1973).

\bibitem{BABAR_jpsiks}
  B.~Aubert {\it et al.} (BaBar Collab.), Phys. Rev. Lett. {\bf 89}, 201802 (2002).

\bibitem{Belle_jpsiks}
  K.~Abe {\it et al.} (Belle Collab.), Phys. Rev. D {\bf 66}, 071102 (2002).

\bibitem{dspdsmks}
  T.~E.~Browder, A.~Datta, P.~J.~O'Donnell and S.~Pakvasa, Phys. Rev. D {\bf 61}, 054009 (2000).

\bibitem{Sanda}
A.~B.~Carter and A.~I.~Sanda, Phys. Rev. Lett. {\bf 45}, 952 (1980); 
A.~B.~Carter and A.~I.~Sanda, Phys. Rev. D {\bf 23}, 1567 (1981); 
I.~I.~Bigi and A.~I.~Sanda, Nucl. Phys. {\bf 193}, 85 (1981).

\bibitem{KEKB}
  S.~Kurokawa and E.~Kikutani, Nucl. Instr. and Meth. A {\bf 499}, 1 (2003),
  and other papers included in this volume.

\bibitem{Belle}
  A.~Abashian {\it et al.} (Belle Collab.),
  Nucl. Instr. and Meth. A {\bf 479}, 117 (2002).

\bibitem{Ushiroda}
  Z.~Natkaniec {\it et al.} (Belle SVD2 Group), Nucl. Instr. and Meth. A {\bf 560}, 1(2006).

\bibitem{SFW}
  G.~C.~Fox and S.~Wolfram, Phys. Rev. Lett. {\bf 41}, 1581 (1978).

\bibitem{res_func}
  H.~Tajima {\it et al.}, Nucl. Instr. and Meth. A {\bf 533}, 370 (2004). 

\bibitem{EvtGen}
  D.~J.~Lange, Nucl. Instr. and Meth. A {\bf 462}, 152 (2001). 

\bibitem{GEANT}
  R.~Brun {\it et al.}, GEANT 3.21, CERN DD/EE/84-1 (1984).

\bibitem{ARGUS}
  H.~Albrecht {\it et al.} (ARGUS Collab.), Z. Phys. C {\bf 48}, 543 (1990).

\bibitem{PDG}
  W.-M.~Yao {\it et al.}, Journal of Physics G {\bf 33}, 1 (2006)

\bibitem{TaggingNIM}
  H.~Kakuno {\it et al.}, Nucl. Instr. and Meth. A {\bf 533}, 516 (2004). 

\bibitem{TSI}
  O.~Long, M.~Baak, R.~N.~Cahn and D.~Kirkby, Phys. Rev. D {\bf 68}, 034010 (2003).

\bibitem{TSI2}
  K.~F.~Chen {\it et al.}, Phys. Rev. D {\bf 72}, 012004 (2005).

\bibitem{BABAR_dspdsmks}
  B.~Aubert {\it et al.} (BaBar Collab.), Phys. Rev. D {\bf 74}, 091101(R) (2006).

\end{thebibliography}
\end{document}